# Modelling Complexity for Policy: Opportunities and Challenges

Bruce Edmonds, Manchester Metropolitan University
bruce.edmonds@gmail.com
Carlos Gershenson, Universidad Nacional Autónoma de México
cgg@unam.mx

This chapter reviews the purpose and use of models from the field of complex systems and, in particular, the implications of trying to use models to understand or make decisions within complex situations, such as policy makers usually face. A discussion of the different dimensions one can formalise situations, the different purposes for models and the different kinds of relationship they can have with the policy making process, is followed by an examination of the compromises forced by the complexity of the target issues. Several modelling approaches from complexity science are briefly described, with notes as to their abilities and limitations. These approaches include system dynamics, network theory, information theory, cellular automata, and agent-based modelling. Some examples of policy models are presented and discussed in the context of the previous analysis. Finally we conclude by outlining some of the major pitfalls facing those wishing to use such models for policy evaluation.

## Introduction

For policy and decision-making, models can be an essential component, as models allow the description of a situation, the exploration of future scenarios, the valuation of different outcomes and the establishment of possible explanations for what is observed. The principle problem with this is the sheer complexity of what is being modelled. A response to this is to use more expressive modelling approaches, drawn from the "sciences of complexity"—use more complex models to try and get a hold on the complexity we face. However, this approach has potential pitfalls as well as opportunities, and it is these that this chapter will attempt to make clear. Thus, we hope to show that more complex modelling approaches can be useful, but also to help people "fooling themselves" in the process.

The chapter starts with an examination of the different kinds of model that exist, so that these kinds might be clearly distinguished and not confused. A section follows on the kinds of uses to which such models can be put. Then we look at some of the consequences of the fact that what we are modelling is complex and the kinds of compromises this forces us into, followed by some examples of models applied to policy issues. We conclude by summarising some of the key danger and opportunities for using complex modelling for policy analysis.

**Kinds of Model**

A *model* is an abstraction of a phenomenon. A *useful model* has to be simpler than the phenomenon but to capture the relevant aspects of it. However, what needs to be represented in a model and what can be safely left out is often a matter of great subtlety. Since relevance changes with context, some models will be useful in some circumstances and useless in others. Also, a model that is useful for one purpose may well be useless for another. Many of the problems associated with the use of models to aid the formulation and steering of policy derive from an assumption that a model will have value for a purpose or in a context different form the one the model was established and validated for. In other words, the value of a model is seen to be in its representation

(e.g. simulation code) and not in its social embedding.

Generally speaking, all of epistemology deals with models. That is to say that all our descriptions can be thought of as models: they are abstractions of the phenomena with which we deal. These abstractions are required in order to understand and communicate about the complex phenomena that we have to deal with. However, models, in this most general sense, are not necessarily either precise or formal. Indeed, most of the models we use in everyday life are informal and couched in a language that is open to a considerable degree of interpretation. Two dimensions of formality can be distinguished:

a) the extent to which the referents of the representation are constrained, e.g. by definition ("specificity of reference" or $SR$),
b) the extent to which the ways in which instantiations of the representation can be manipulated are constrained, e.g. by rules of logical deduction ("specificity of manipulation" or $SM$).

These two dimensions are illustrated in figure 1, below.

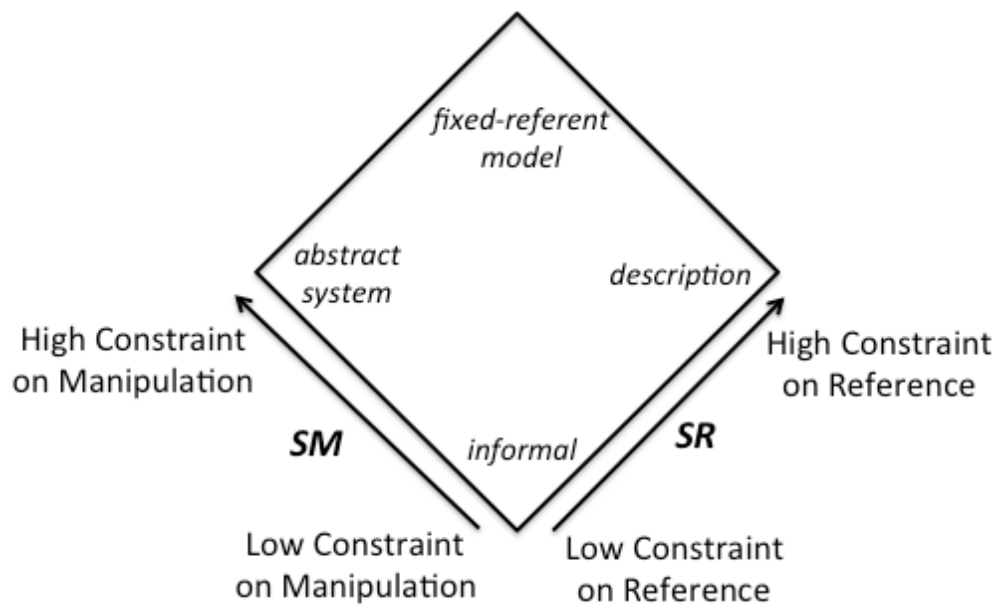

**Figure 1. Two dimensions of formality**

For example, an analogy expressed in natural language has a low $SR$ since, what its parts refer to are reconstructed by each hearer in each situation. For example, the phrase "*a tidal wave of crime*" implies that concerted and highly coordinated action is needed in order to prevent people being engulfed, but the level of danger and what (if anything) is necessary to do must be determined by each listener. In contrast to this is a detailed description where what it refers to is severely limited by its content, e.g. "*Recorded burglaries in London rose by 15% compared to the previous year*".

A system of abstract logic, mathematics or computer code has high SM since the ways these can be manipulated is determined by precise rules—what one person infers from them can be exactly replicated by another. This in contrast to a piece of natural language which can be used to draw inferences in many different ways, only limited by the manipulators' imagination and linguistic ability. However, just because a representation has high SM does not mean that the meaning of its parts in terms of what it represents is well determined. Many simulations, for example, do not represent anything we observe directly, but are rather explorations of ideas. We, as intelligent interpreters, may mentally fill in what it might refer to in any particular context, but these "mappings" to reality are not well defined. Such models are more in the nature of an analogy, albeit one in formal form – they are not testable in a scientific manner since it is not clear as to precisely what they represent. Thus simulations, especially agent-based simulations, can give a false impression of their applicability because they are readily interpretable (but informally). This does not mean they are useless for all purposes. For example, Schelling's abstract simulation of racial segregation did not have any direct referents in terms of anything measurable[1], but it was an effective

---

[1] Subsequent elaborations of this model have tried to make the relationship to what is observed

counter-example that can show that an assumption that segregation must be caused by strong racial prejudice was unsound. Thus such 'analogical models' (those with low $SR$) can give useful insights, they can inform thought, but they can not give reliable forecasts or explanations as to what is observed.

Formal models are a key aspect of science, since scientific models aim at describing and understanding phenomena. Their formality is important because that means that both their inference and meaning is (a) checkable by others and (b) stable. This makes it possible for a *community* of researchers and others to work collectively with the same models, confident that they are not each interpreting them in different ways. As the above discussion should have made clear, they can be formal in (at least) two different ways. For example data is a formal model of some aspect of what we observe, in the sense that it abstracts but in a well defined way – its meaning is precise. Data is not formal in terms of $SM$ however, and one could make very different inferences from the same set of data. Usually "formal modelling" means that the inference from a model is well specified, in other words it is a representation with high $SM$. Thus scientific modelling is often associated with mathematics or computer simulation. However, in order to connect the formal inference to data it has to be formal in the $SR$ sense as well, there needs to be a precise mapping between its parts and processes to what is observed, which is usually[2] done in terms of a map to some data.

**The Use of Models**

There are many purposes for models, including: as a game, an

---

more direct, but the original model, however visually suggestive, was not related to any data.
[2] It is possible to directly 'wire' something like a computational process to reality via sensors and actuators, as happens in programmed trading, in this case it is not always clear the extent to which the model is a representation of anything observed, but more an embedded participant in it.

aesthetic construction, or an illustration of some idea[3]. Most scientific models claim to be *predictive*, i.e. they should allow us to obtain information about the future of the phenomenon before it occurs. For example, one can calculate and predict a ballistic trajectory aiming at a target using a mechanical model. However, on closer examination, many are more concerned with two other goals: *explanation* or *exploration*, with that of prediction being left as a theoretical possibility only[4]. There are many scientific models that are not predictive, or which only predict abstract properties. For example, the Gutenberg–Richter law describes the distribution of earthquake intensities, but this does not tell us when might be the next earthquake nor how intense it might be. Darwin's theory of evolution does not tell us what will evolve next, or even the reasons why what has evolved did so, but it does predict the relationship between genetic distance and the length of time since species diverged. Unfortunately many reports about models are not clear as to their purpose in this regard, indeed many seem to deliberately conflate different purposes. Whilst models may have more than one goal, one should be wary of a model that was developed and tested for one purpose but is now being used for another. A clear case of this is where a model is designed to establish a theoretical counter-example (such as in the Schelling case discussed below) but then is later claimed to be for prediction (albeit in a modified form).

There is another, very basic, distinction in the way models are used in practice. That is between models that (a) represent something observed and ones which (b) are a component of an adaptive strategy.

---

[3] (Epstein, 2008) lists 16 different reasons, (Edmonds, et al. 2013) considers reasons that are more connected with understanding human society.
[4] It is common for papers describing them to list prediction as "future work" when the model is more fully developed.

In the former case, there is a well-defined mapping between the model and observational data/measurements, and the model is judged as to the extent of its error in its predictions of its target phenomena. Here the model is, to different degrees and ways, either correct or not. In this case an examination of the model can tell us something about the structure of what is modelled, for example by exploring "what-if" questions using the model. For sake of clarity we call this a "representational" model.

In the later case, the model is part of a decision making process to select strategies for action. It takes (processed) inputs from the world, for example indicators of success and the model is changed depending on how well it is doing (for example by depreciating the parts of the model that resulted in a poor indicator). Outputs from the model are used in the determination of interventions. Here the model is continually being adapted according to events, it somehow encodes past successes or failures for different courses of action given different observations of the world. Here a useful model may not represent any aspect of the world at all, but just be a useful intermediary in the process of decision making. However, if the process of adaption is effective it may come to encode knowledge as to what works and what does not. We call this an "adaptive" model.

It may well be that an adaptive use of a model is more effective in a particular setting, particular if a considerable period of adaption has occurred, in effect training the model (given the decision making structure it is embedded in) using a considerable amount of feedback from its policy environment. If the model is sufficiently flexible (i.e. has many adjustable internal parameters) that it could indicate the correct action from the available inputs (derived from observations of the environment) then, with enough training, the model will eventually do so. However, this kind of model adaption means that it is probably finely tuned to the particular situation and will not be useful by others in similar situations. Nor is it likely to be much use in exploring what would happen in cases not yet observed; so if the situation changes in some fundamental way, the model may well give totally the wrong answers. Furthermore, it might not be apparent from an inspection of the adapted model, why it works.

Representational models are usually hard to develop, taking considerable time and effort, often by a team of experts somewhat separate from those making policy decisions. Such models usually rely on some theory of the system being modelled, whose assumptions may be explicit. This kind of model, if it validates well, might have some validity outside its original test situation, and moreover, its assumptions and structure might give clues as to when and how it might reliably be used. If the situation changes in a way that is explicitly encoded into the model, one might be able to change its settings to suit the new situation.

In practice, models are often used with a mixture of adaptive and representational models, with adaptive models encapsulating some theory and being somewhat representational, and representational models undergoing some process of model adaption over time. In this case it is wise to know which aspects of ones model are representational and which have been 'tuned' to the particular situation or set of data.

Models are limited and using them carelessly can have counterproductive consequences. One of the main limitations is due to the *complexity* of their subject matter, which is what we discuss next.

## Complexity and its Implications for Modelling

That society is complex may seem an obvious statement. Still, the ways in which it is complex has implications for the use of models for the planning and execution of policy. One problem is the lack of agreement as to what "being complex" means. There are dozens different definitions of complexity (Edmonds, 2000). Frequently the word is used as a kind of negative. When available techniques (or accepted techniques) fail we call what were trying to analyse, "complex"—in this case it is a "dustbin concept", a category to use when others fail. Here, in order to obtain a common understanding of the term, we can use its etymology. Complexity comes from the Latin *plexus*, which means interwoven. Thus something complex is difficult to separate out into separate components or processes. This is because of *relevant interactions* (Gershenson and Heylighen, 2005; Gershenson, 2013a). Interactions are relevant when the future of an element of a system is partially determined by the interactions, in other words, if one eliminates these interactions then the future would be significantly different.

Traditionally, models have been reductionist, in the sense that they study phenomena in isolation. By definition, interactions are excluded. Either an element is modelled in isolation, or a whole system is modelled, averaging the properties of its elements. In other words, traditionally phenomena are modelled at a single scale. This approach is suitable for simple systems, but it is not sufficient for complex ones, where the properties of the system are a consequence of the interactions of the elements. This requires models to be multi-scale (Bar-Yam, 2004), and interactions must be modelled to relate different scales.

In terms of policy models, interactions need to be included in the model if the future projections are not to be distorted. Simple models that do not include interactions will be unreliable. Such interactions carry important implications for modelling complex systems. These all make the ideal of the assessment of the impact of policy interventions using a model difficult.

*Firstly*, it implies that elements cannot be studied in isolation.

Different social processes can interact to produce effects different from those caused by each process singly. The outcome from a population that is both disaffected and has access to an effective medium for dissemination of views (e.g. twitter) might well be very different to that of a disaffected population with only local gossip or a satisfied population with something like twitter.  The impact of this is that separately analysing the impacts of different factors upon the outcome might well be misleading; one has to consider the outcomes from the whole system.  This leads to the problem that one might well not know how much one needs to include in an model adequate for ones purposes, and that an approach that starts with the simplest possible model and then experimentally adds processes one at a time, might never get you to an adequate model (Edmonds & Moss, 2005).

*Secondly*, it implies that interactions generate *novel information*, which is not present in initial or boundary conditions. This new information inherently limits the predictability of a complex system. In other words, the results are at least partially 'caused' by processes within the system, and not by external factors that can be controlled for.  At best, this may mean that one has to make do with a broad distribution of outcomes as a forecast, or the prediction of 'weaker', second-order properties of the outcome (e.g. the volatility of the focus outcomes, or what will *not* happen). At worst, it may mean that there is no well-defined distribution of outcomes at all, with any measures upon the outcomes from a model being due to artefacts (e.g. model size).

The result of such difficulties means that any policy model is inevitably a compromise between different desirable modelling goals.  Figure 2 below illustrates some of these tensions in a simple way.

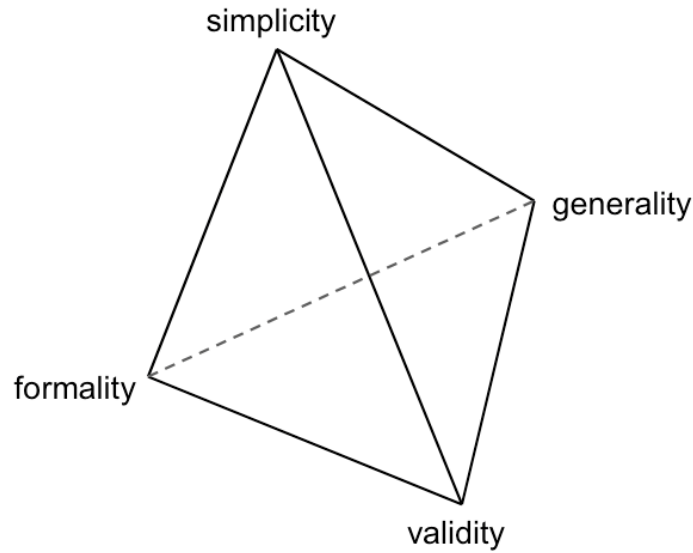

**Figure 2. Some of the tensions implicit in modelling complex systems.**

These illustrated desiderata all refer to the model that is being used. *Simplicity* is how simple the model is, the extent to which the model itself can be completely understood. Analytically solvable mathematical models, most statistical models and abstract simulation models are at the relatively simple end of the spectrum. Clearly a simple model has many advantages in terms of using the model, checking it for bugs and mistakes (Galan, et al. 2009) and communicating it. However, when modelling complex systems, such as those policy makers face, such simplicity may not be worth it if gaining it means a loss of other desirable properties. *Generality* is the extent of the model scope: how many different kinds of situation could the model be usefully applied. Clearly *some* level of generality is desirable; otherwise one could only apply the model in a single situation. Authors are often rather lax about making the scope of their models clear—often implying a greater level of generality that can be substantiated. *Formality* is what was called specificity of reference (SR) above. Models where the meaning of the model parts are well-defined have such formality, those which do not, and are more in the way of a model of ideas about some target system, rather than the system directly have less of this.

Finally *validity* means the extent to which the model outcomes match what is observed to occur—it is what is established in the process of model validation. This might be as close a match as a point forecast, or as loose as projecting qualitative aspects of possible outcomes.

What policy makers want, above all, is validity, with generality (so they do not have to keep going back to the modellers) and simplicity (so there is an accessible narrative to build support for any associated policy) coming after this. Formality is for them is not a virtue but more of a problem, they may be convinced it is necessary (so as to provide the backing of 'science'), but it means that the model is inevitably somewhat opaque to them and not entirely under their control. Modellers, usually, have very different priorities. Formality is very important to them so that they can replicate their results and so that the model can be unambiguously passed to other researchers for examination, critique and further development (Edmonds 2000). Simplicity and generality are nice if you can get them, but one cannot assume that these are achievable (Edmonds 2012). Validity *should* be an overwhelming priority for modellers; otherwise they are not doing any sort of empirical science. However, they often put this off into the future, preferring the attractions of the apparent generality offered by analogical models (Edmonds 2001, Edmonds 2012). Relatively simple models that explore ideas rather than relate to any observed data that give the illusion of generality are, unfortunately, common.

Another ramification of the complexity of what is being modelled is in the goal of modelling—what sort of purpose the model suitable for. One of the consequences of this is that prediction of policy matters is hard, rare, and only obtained as a result of the most specific and pragmatic kind of modelling developed over relatively long periods of time[5]. It is more likely that a model is appropriate for establishing and understanding candidate explanations of what is happening, which will inform policy making in a less exact manner than prediction, being part of the mix of factors that a policy maker will take into account when deciding action. It is common for policy people to want a prediction of the impact of possible interventions "however rough", rather than settle for some level of understanding of what is happening, however this can be illusory. If one really wanted a prediction "however rough" one would settle for a random prediction[6] dressed up as a complicated "black box" model. If we are wiser, we should accept the complexity of what we are dealing and reject models that give us ill-founded predictions.

One feature of complex systems is that they can result in completely unexpected outcomes, where due to the relevant interactions in the system, a new *kind* of process has developed resulting in qualitatively different results. It is for this reason that complex models of these systems do not give probabilities (since these may be meaningless, or worse be downright misleading) but rather trace some (but not all) of the possible outcomes. This is useful as one can then be as prepared as possible for such outcomes, which otherwise would not have been thought of.

The effective use of models for policy formulation will thus involve a clear focus as to its purpose and its manner of use combined with some compromise between the factors discussed above. However, the extent and impact of such compromises should be openly and honestly made, as a proper balance is necessary for reliable uses of the model. It is probable that a combination of related models, each making different

---

[5] For an account of actual forecasting and its reality, see Silver (2012).
[6] Or other null model, such as "what happened last time" or "no change".

compromises might be a productive way forward. However, this requires extra work and care.

We now look at a number of different approaches, commenting upon the compromises and properties of each.

## Tools and Approaches

### *System Dynamics*

System dynamics is an approach to modelling that represents a system in terms of a set of interconnected feedback loops (Forrester, 1971). It models these in terms of a series of flows between stocks plus additional connections between variables and flows. Crucially, it allows the representation of delays in such feedback and that the outcomes of some variables can control/effect the rate of other flows. These flows and relationships can then be simulated on a computer and (more recently) visualized. Its advantages are that a complex set of feedback relationships can be explored and hence better understood. However, in practice, the variables it deals with are themselves abstract entities, often representing abstract and aggregate quantities. This approach is not well suited to the modelling of systems where internal heterogeneity is significant in terms of determining the outcomes.

### *Network Theory*

Networks naturally describe complex systems, representing elements as nodes and their interactions explicitly as links. Only in the last decade, there has been an explosion in the scientific exploration of networks and their application to a broad range of domains. Network theory has its roots in graph theory as proposed by Euler in the eighteenth century. However, it is only recently that its use has become widespread, in part because of the large computing power and big data sets available.

Networks are useful for representing the *structure* of systems, indicating how elements interact. However, they can also represent the *function* of systems, with nodes representing states and directed links representing transitions. Relating the structure and function of systems is one of the most common questions for understanding systems, i.e. how changes in the structure affect the function of a system? Network theory can be used to study both structure and function using the same formalism. Also, adaptive and temporal networks (Gross & Sayama, 2009; Holme & Saramäki, 2012) have been used to study the change in time of network structure.

From the study of different natural and artificial networks, it has been found that most of them do not have a trivial topology, i.e. there is a relevant organization in their structure. Still, several modelling approaches assume homogeneous topologies, as in cellular automata (see below), or even a so called "well mixed" population, i.e. there is no structure considered (only the macro state). It has been shown that structure (micro scale) plays a crucial role in the dynamics of such systems. For example, the same system may change drastically its dynamics depending on whether the local structure is considered or not (Shnerb et al., 2000).

Network models can be useful to study several aspects related to policy and decision making. For example, random agent networks (RANs) were proposed to model organizations such as bureaucracies (Gershenson, 2008), showing how few modifications to the structure of an organization can improve considerably its performance. In general, "computing networks" (Gershenson, 2010) can be used to study and relate adaptability at different scales. Since policy and decisions are usually made over changing and uncertain scenarios, adaptability is a desired property of models. However, the more networks change and the complexity of the interactions represented over the links get, the less classic network theory is applicable, and the closer to an individual-based model one has.

***Information theory***

Claude Shannon (1948) proposed information theory in the context of telecommunications. He was interested on how a message could be transmitted reliably over unreliable media. He proposed a measure of information (equivalent to the Boltzmann-Gibbs entropy in thermodynamics) where information is minimal for regular strings, as new symbols do not carry new information. Shannon information is maximal for random strings, as new symbols carry all the new information, i.e. they are not predictable. Several other measures have been derived from Shannon's information, such as mutual information, predictive information, excess entropy, and information transfer, among others (Prokopenko et al., 2009).

Information theory has been used repeatedly to measure complexity. However, there are two different views. One view implies a similarity of information to complexity, where maximum randomness (Shannon information) would have maximum complexity. A more popular view poses that complexity is maximal when a balance between regularity (order) and randomness (chaos) is reached (Langton, 1990; Kauffman, 1993).

Recently, measures of complexity, emergence, self-organization, homeostasis, and autopoiesis were proposed based on information theory (Fernández et al., 2014). These measures are fast to compute and simple enough to be used by people without a strong mathematical background, but can give insights into the dynamics of systems. It has been argued (Edmonds 1999) that there is not one such measure that can always be used, but rather one has to choose a measure that gives meaningful results for the kind of system that one is considering.  Thus this approach assumes that one has understood the target system sufficiently to select the appropriate measure.

For decision making, it is vital to identify which type of dynamics are followed by systems and their components, as different decisions should be made depending on regular, complex, or chaotic dynamics. Used correctly, these measures can provide precisely this information and thus aid in knowing how to respond to change in the systems.

### *Cellular automata*

Cellular automata (CA) can be seen as a particular type of network. Each cell (node) has a state that depends on the states of its neighbours (links) and its own previous state. Different CA models can have different number of states and consider different number of neighbours. Cells can also be arranged in one dimension (array), two dimensions (lattice), three, or more dimensions.

Perhaps the most popular CA is Conway's "Game of Life" (Berkelamp, et al., 1982). Each cell can have one of two states: '0' (dead) or '1' (alive). Rules consider how many of the eight closest neighbours are alive. For a live cell to continue living, it must have two or three living neighbours (in any configuration). More than three or less than two neighbours implies that the cell will die in the next time step. New cells are born on empty cells when they have exactly three neighbours. With these simple rules, several complex structures emerge: stable structures of different shapes, oscillators of different periods, moving structures (gliders, spaceships), eaters, glider guns, etc. The structures emerging with the simple rules of the Game of Life can be used even to build a Universal Turing Machine[7]. An example of the dynamics of the Game of Life is shown in Figure 3.

---

[7] To explore the Game of Life and other interesting CA, the reader is advised to download Golly at http://golly.sourceforge.net

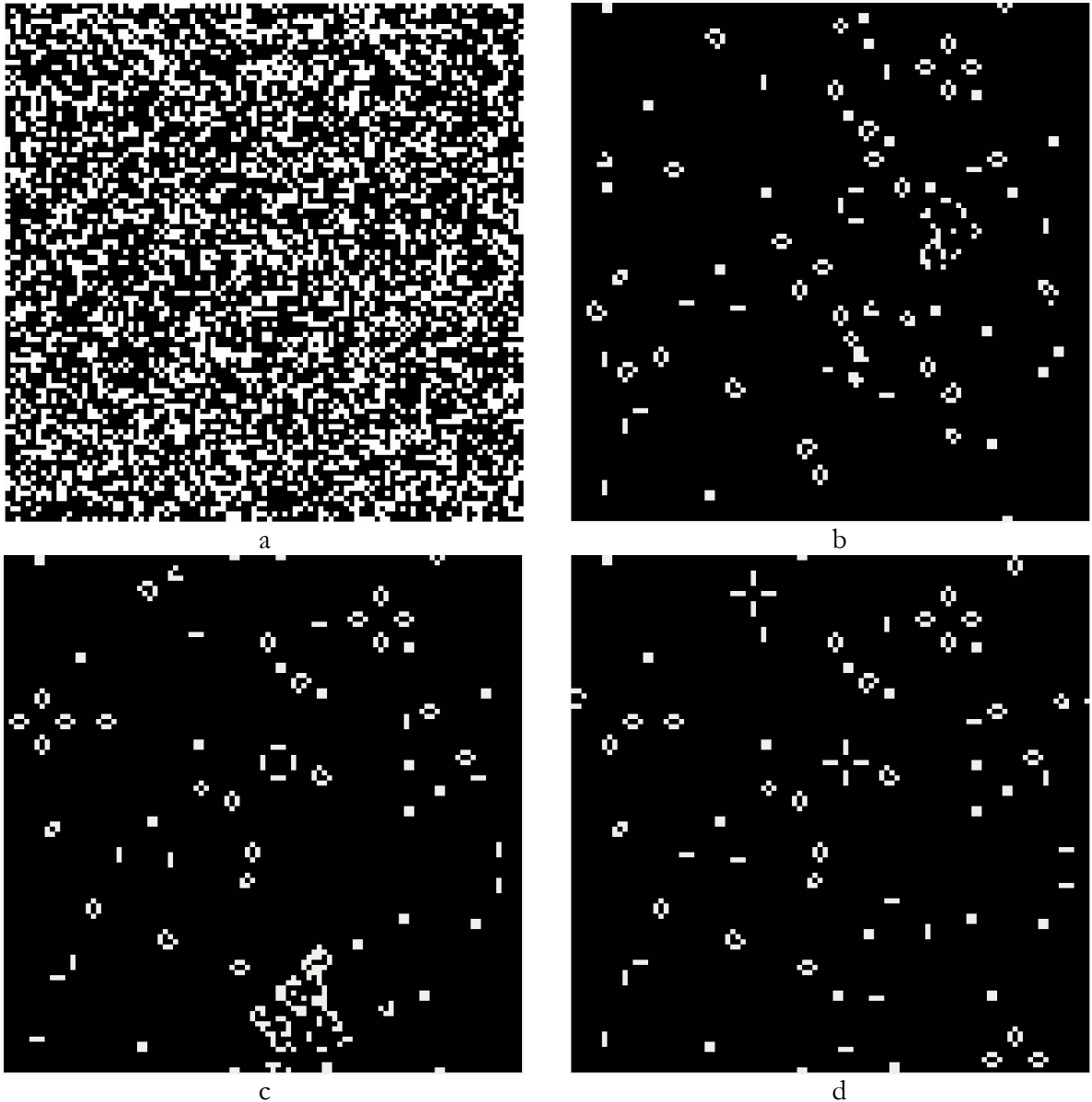

Figure 3. Evolution of the Game of Life from a random initial condition (a), where white cells are "alive" and black cells are "dead". After 410 steps (b), certain stable structures have been formed, but there are still some active zones. After 861 steps (c), some structures have been destroyed and some new ones have been created. Activity continues in the lower part of the lattice. After 1416 steps (d), the dynamics is periodic, with stable and oscillatory structures. Images created with NetLogo (Wilensky, 1999). Figure initially published in Gershenson (2013b).

Cellular automata have been used in several urban and land-use models (e.g. Portugali, 2000; Batty, 2005). However, "pure" CA models tend to be too formal and abstract, so they have been found to be more useful in hybrid models, in many cases combining CA with agent-based modelling.

### *Individidual- and Agent-based modelling*

Individual-based modelling is given when social actors or entities are represented by separate 'objects' within a computational simulation. Each object can have different properties, so this technique can represent heterogeneous collections of individuals. The interactions between the actors are represented by messages between the objects of the simulation. Thus, the mapping between what is observed and the model can be very much more straightforward with such simulations: each object modelling its corresponding actor.

When the objects in the simulation have internal processes representing their learning or decision making processes so that these processes could be usefully interpreted as cognition, we call the computational objects "agents" since they can act somewhat independently—they have a simple form of agency. When the agents are of this form, one has the technique of Agent-Based Modelling (ABM). This technique is very flexible and puts few constraints upon the modeller, so the simulations that result are difficult to characterize in general but are of various kinds. An accessible introduction to the approach is (Gilbert & Troitzsch, 2005) and a more comprehensive guide (Edmonds & Meyer, 2013).

In particular, simulations differ greatly as to their level of detail, ranging from highly abstract and relatively simple simulations to very specific and complicated ones. The key difference here is whether the driver for model development is simplicity or relevance (Edmonds & Moss, 2005). Agent-based modelling has now been applied to a large number of policy-relevant subjects, including (to take an arbitrary sample of recent applications): energy infrastructure siting (Abdollahian, et al., 2013), password behaviours within an organization (Renaud, et al., 2013), mobile banking adoption (Wei, et al., 2013), China's housing market (Zhang, et al., 2013) and return migration (Biondo, et al. 2013).

The problem with ABM lies not in its expressiveness but in the complexity of its models (which means that it may be hard to understand the models themselves) and establishing the relationship between the models and what they represent (Moss & Edmonds 2005).

## Examples

### *Club of Rome's "Limits to Growth"*

In the early 1970's, on behalf of an international group under the name "The Club of Rome" a simulation study was published (Meadows et al. 1972) with the attempt to convince humankind that there were some serious issues facing it, in terms of a coming population, resource, and pollution catastrophe. To do this they developed a system dynamics model of the world. Thus, this is a fairly simple kind of model that does not explicitly represent the parts of a system or its interactions, but rather the feedback cycles between key global factors. It was important to the authors to go beyond simple statistical projections of the available data, since that missed out the crucial delays in the usually self-correcting feedback processes. The results of the simulations were a set of computed curves showing such as pollution, population, etc. The results indicated that there was a coming critical point in time and that a lot of suffering would result, even if humankind managed to survive it.

The book had a considerable impact, firmly establishing the idea that it was possible that humankind could not continue to grow indefinitely. The book presented the results of the simulations as predictions—a series of what-if scenarios. Whilst they did add caveats and explore various possible versions of their model, the overall intent of the book was unmistakable: that if we did not change our lifestyles, disaster would result.

The authors clearly hoped that by using a simulation they would be able to make the potential feedback loops real to people. Thus this was a use of simulation to illustrate an understanding that the authors had. It was thus a model of ideas rather than directly of any such data. It did not, and could not, make predictions about what will happen in the future, but rather illustrate some possibilities. However, the model was not presented as such, but as something more scientific in some sense. It was the presentation as 'scientific' that made this book such a challenge but also what laid it open to criticism (e.g. Cole, et al., 1973). An examination of the model showed that some of its parameters were very sensitive and thus had to be 'tuned' to get the published results (Vermeulen & de Jongh, 1976). In other words, whilst the models had an illustrative and exploratory purpose, they were presented and criticised as if it was a predictive model.

The book made a considerable impact upon the general consciousness of the problem, and did act to get people questioning previously held assumptions (that we could keep on growing economically and physically). However, it was also largely discredited in the eyes of other modellers due to its perceived lack of 'rigor'. This was somewhat unfair as the alternatives were no better in terms of validity or generality. However, a lack of humility in terms of its results and the relative simplicity of their model did lay it open to such attacks. The predictions of the book have not yet come to pass, but it is not clear that a similar future critical point and attendant suffering has been avoided.

*Schelling's Model of Racial Segregation*

In addition to a host of simpler, analytically expressed models (similar in kind to the Club of Rome's systems dynamics models), Schelling developed what we might recognize as a simple agent-based model (Schelling, 1971; 1978). It did represent individuals and their neighbourhoods explicitly, albeit abstractly in a 2D grid with black and white 'counters' representing the people. The simulation was very simple—counters were distributed randomly to start with then each counter that had less than a given percentage ($c$) of like neighbours moved to a new empty spot. The simulation showed that segregation emerged even with relatively low levels of racial bias (values of $c$ down to 30%). This did not relate to any particular data but was rather a counter-example to the idea that the observed segregation must be due to strong racial prejudice. In other words this was intended to be an exploration of ideas to inform policy rather than a direct representation of what was happening. It produced an understanding of possible segregation processes, and so influenced local policies in Chicago, away from focusing on prejudice as a cause of the extreme segregation they had.

*Employment in an Arctic Community*

Berman et al. (2004) consider eight employment scenarios defined by different policies for tourism and government spending, as well as different climate futures, for an ABM case study of sustainability in the Arctic community of Old Crow, in Canada. Scenarios were developed with the input of local residents: tourism being a policy option largely influenced by the autonomous community of Old Crow (stemming from their land rights), and attracting great local interest. Here the policy options were addressed as a certain type of scenario, embedding the behaviour of actors within a few possible future contexts. The simulation here ensured the consistency of the scenarios, and helped to integrate the various inputs into a coherent whole.

The merit of this model is that it can improve the reckoning of human and social factors and information into the issues at stake; allowing the exploration of some real possible outcomes. The drawback is the multiplication of uncertainties, not least of which is that we do not convincingly know how social actors might adapt to new circumstances (even if the policy options are relatively concrete).

### *A detailed model of HIV spread and social structure*

Alam et al. (2007) investigate the outcomes indicated by a complex, and detailed model of a particular village in the Limpopo valley of South Africa. This model in particular looks at many aspects of the situation, including: social network, family structure, sexual network, HIV spread, death, birth, savings clubs, government grants and local employment prospects. It concludes with hypotheses about this particular case, showing that complex destructive synergies between the spread of HIV and the breakdown of social structure were possible, and could be exacerbated by the influx of workers from outside due to the granting of mining concessions. This does not mean that these outcomes will actually occur, but this does provide a focus for future field research and may provide thought for policy makers. Unfortunately in this case, the conclusions of this study were not what the local authority wanted to hear, and so the findings were ignored. This was a model with a high degree of validation, but a very specific and complex model taking 3 years to develop.

### *Evaluating Pandemic Preventive Measures*

Bajardi et al. (2011) combined in a model networks and agents to model the epidemic spread of the H1N1 influenza in 2009. Nodes represent regions, which are linked by the commercial flights between their major airports. At each node, agents can be susceptible, exposed, infected, or recovered (Anderson & May, 1992). Adjusting different parameters, the global spread of the disease could be reproduced. Travel restrictions were imposed as a preventive measure; reducing air travels by 40%. Simulations showed that travel restrictions are ineffective to prevent the spread of the disease. Comparing with scenarios with no travel restrictions or with even more stringent travel restrictions, the authors found that the disease reached a peak almost on the same day.

This is a theory-based model intended for predictive purposes. Its validity depends upon the approximations and assumptions in the model, including the characteristics of the social network used in the model.

**Prospects and Dangers for Complex Policy Modelling**

All models have limits—not only limits in the accuracy of their predictions but in the expression of the situations under which such projections are based. The nature of complex situations means that attempts to use models to aid policy formulation are susceptible to some particular dangers and pitfalls, including:

- Confusing a model that has exploratory or explanatory purposes for one that is predictive.
- Preferring a 'black box' model that seems to give definite predictions despite neither understanding it nor knowing that it is reliable for this purpose.
- Trying to use a model that has been adapted within a highly specific situation out of its original context.
- Attempting very general or simple models of policy issues probably means sacrificing direct validity for an indirect, analogical relationship only.

Complex models of the kind described here (and elsewhere in this book) have the potential to express a broader range of kinds of situations than previous approaches. They are thus not so limited by the kind of 'brave' assumptions that bedevil models where analytic results are deemed necessary. They are also ideal for the exploration and "laying bare" complex dynamics. For this reason, they are prospective as an important tool in the exploration and consideration of policy options. In particular, more descriptive models can be directly related to what they are modelling, allowing a greater range of data and input to be utilized in their specification and validation (both high SM and SR in terms of the above discussion).

Given the fact that models of complex systems will offer a limited predictability, it is advisable to complement this lack of predictability with adaptability (Gershenson, 2007). This will enable decision makers to take the best choice for the specific circumstances that are faced, as adaptability implies a distinction of current circumstances that purely predictive models do not consider.

## Acknowledgements

Bruce Edmonds acknowledges the support of the EPSRC under grant number EP/H02171X/1. Carlos Gershenson was partially supported by SNI membership 47907 of CONACyT, Mexico.